# Strong enhancement of chlorophyll *a* concentration by a weak typhoon

Liang SUN[1,2], Yuan-Jian Yang[3], Tao Xian[1], Zhu-min Lu[4] and Yun-Fei Fu[1]

1. Laboratory of Atmospheric Observation and Climatological Environment, School of Earth and Space Sciences, University of Science and Technology of China, Hefei, Anhui, 230026, China;

2. LASG, Institute of Atmospheric Physics, Chinese Academy of Sciences, Beijing 100029, China

3. Anhui Institute of Meteorological Sciences, Hefei 230031, China;

4. Key Laboratory of Tropical Marine Environmental Dynamics, South China Sea Institute of Oceanology, Chinese Academy of Sciences, Guangzhou 510301, China.

Corresponding author address:

Liang SUN

School of Earth and Space Sciences, University of Science and Technology of China

Hefei, Anhui 230026, China

Phone: 86-551-3606723;

Email: sunl@ustc.edu.cn; sunl@ustc.edu

**Abstract:**

Recent studies demonstrate that chlorophyll *a* (chl *a*) concentrations in the surface ocean can be significantly enhanced due to typhoons. The present study investigated chl *a* concentrations in the middle of the South China Sea (SCS) from 1997–2007. Only the Category1 (minimal) Typhoon Hagibis (2007) had a notable effect on the chl *a* concentrations. Typhoon Hagibis had a strong upwelling potential due to its location near the equator, and the forcing time of the typhoon (>82 h) was much longer than the geostrophic adjustment time (~63 h). The higher upwelling velocity and the longer forcing time increased the depth of the mixed-layer, which consequently induced a strong phytoplankton bloom that accounted for about 30% of the total annual chl *a* concentration in the middle of the SCS. The implication is that the forcing time of a typhoon should be long enough to establish a strong upwelling and consequently for the induction of significant upper ocean responses.

## 1. Introduction

Investigations of typhoons' impacts on the upper ocean have concluded that the surface chlorophyll-a (Chl-a) concentrations substantially increase after passage of a typhoon [McClain et al. 2004, McClain 2009, Subrahmanyam et al. 2002, Lin et al. 2003, Babin et al. 2004, Zheng & Tang 2007, Gierach & Subrahmanyam 2008], especially in the oligotrophic waters [Babin et al. 2004]. The physical mechanism for the increase in Chl-a concentration is that a typhoon's strong wind induces mixing and upwelling in the upper ocean [Price 1981], which brings both subsurface Chl-a to the surface and subsurface nutrients into the euphotic zone [Subrahmanyam et al. 2002, Lin et al. 2003, Babin et al. 2004, Zheng & Tang 2007, Gierach & Subrahmanyam 2008]. Although it is not so clear whether the extra Chl-a is due to upwelling of nutrients or of Chl-a, we are sure that both the typhoon-induced upwelling and the pre-existing eddies favor the Chl-a concentration enhancement [Walker et al. 2005, Shi & Wang 2007, Zheng et al. 2008, McClain 2009, Sun et al. 2009]. The stronger the upwelling is, the more nutrients are taken into the surface. However, whether the typhoons have notable impacts on the ocean primary production is still open to discussion.

On the one hand, the single case studies, where the impacts of the super typhoons were considered, showed that typhoons did have notable impacts on the regional ocean primary production. In the East China Sea, Typhoon Meari induced a three-fold primary production increase, which contributed to 3.8% of the annual new production [Siswanto et al. 2008]. In the South China Sea (SCS), it was also estimated that, on average, 30-fold increases in surface chlorophyll-a concentrations were triggered by Typhoons Kai-Tak (2000) [Lin et al. 2003] and Lingling (2001) [Shang et al. 2008]. The single Typhoon Kai-Tak (2000) and the typhoons from the entire year had induced 2~4% and 20~30% of the SCS's annual new primary production, respectively [Lin et al. 2003].

On the other hand, the integrated impact of tropical cyclones on sea surface Chl-a is contradictory to the above-mentioned point [Hanshaw et al. 2008, Zhao et al. 2008]. The comparative study of Typhoons Lingling (2001) and Kai-Tak (2005) indicated that most of the typhoons in the SCS were relatively weaker compared to Kai-Tak and Lingling, and that typhoons accounted for 3.5% of the annual primary production in the oligotrophic SCS [Zhao et al. 2008]. A similar conclusion was drawn for the North Atlantic that Chl-a concentration contributed to only 1.1% of the positive Chl-a anomaly within the hurricane season, which implies that the integrated impact of tropical cyclones might be ignored [Hanshaw et al. 2008]. Such conclusions are not surprising if we recall the investigation of the hurricane-induced phytoplankton blooms on the Sargasso Sea [Babin et al. 2004]. It was found that 13 hurricanes induced an average of a three-fold (range from one-fold to nine-fold) Chl-a increase for about two weeks within the sea. Thus, each hurricane has induced less than a 1% Chl-a increase toward the annual mean. Considering that annual hurricane numbers are about three to four, the integrated impact of tropical cyclones on Chl-a increase is very small, which is consistent with the integrated estimations [Hanshaw et al. 2008, Zhao et al. 2008].

As the typhoon-induced Chl-a enhancement depends on the amount of subsurface nutrients taken into surface water due to upwelling, the main point of the arguments above lies in whether the typhoon can induce strong enough upwelling. Motivated by such investigations and the above-mentioned arguments, this study investigated the Chl-a concentrations in the South China Sea (SCS) from 1997-2007 and found that only the category-1 Typhoon Hagibis (2007) had notable impacts on the Chl-a concentration.

## 2. Data and methods

The merged daily and monthly Chl-a concentration data (Level 3), with a spatial resolution of 9 km from two ocean color sensors (MODIS and SeaWiFS), were produced and distributed by the NASA Goddard Space Flight Center's Ocean Data Processing System (ODPS).

Typhoon tracking data, taken every six hours, including center location, central pressure, and maximum ten-minute mean sustained wind speeds (MSW), were obtained from the Shanghai Typhoon Institute (STI) of the China Meteorological Administration (CMA). Additionally, two other types of wind data were used. One was the sea surface wind (SSW) vector and stress with a spatial resolution of 1/4°×1/4°, obtained from the daily QuikSCAT (Quick Scatterometer), provided by the Remote Sensing Systems (http://www.remss.com/). The wind stress $\vec{\tau}$ was calculated with the bulk formula [Garratt 1977],

$$\vec{\tau} = \rho_a C_D |\vec{U}| \vec{U}, \tag{1}$$

where $\rho_a$ and $\vec{U}$ are the air density and wind vector, and $C_D = (0.73 + 0.069U) \times 10^{-3}$ is the drag coefficient. According to recent studies [Powell et al. 2003, Jarosz et al. 2007], the above drag coefficient is an overestimation for wind speeds over 40 m/s. The potential upwelling velocity $V_e$ due to wind was calculated by using the Ekman pumping velocity (EPV) formula [Price et al. 1994],

$$V_e = curl(\frac{\vec{\tau}}{\rho f}) \tag{2}$$

where $\rho = 1020$ kg m$^{-3}$ is the density of sea water, $f$ is the Coriolis parameter. The thermocline displacement (or isopycnal displacement) $\Delta \eta$ due to a typhoon with translation speed $U_T$ is estimated as [Price et al. 1994],

$$\Delta \eta = \frac{\vec{\tau}}{\rho f U_T}, \tag{3}$$

The other is the "best track datasets" of the western North Pacific, obtained from the Joint Typhoon Warning Center (JTWC). Each best-track file contains tropical cyclone center locations and intensities (i.e., the maximum one-minute mean sustained ten-meter wind speed) at six-hour intervals. Such maximum wind speeds can be found in Table 1. These one-minute mean sustained wind speeds are relatively larger than ten-minute mean sustained wind speeds, which leads to even larger uncertainty in wind stress calculations, according to recent studies [Powell et al. 2003, Jarosz et al. 2007]. Thus, we used the MSW and a more modern drag law $C_D = (-2.229 + 0.2983U - 0.00468U^2) \times 10^{-3}$ [Jarosz et al. 2007] to calculate the wind stress $\vec{\tau}$. Additionally, wind radius and the radii of specified winds (35, 50, 65 or 100 kts) for four quadrants were also included, which were useful for the wind stress curl calculations. Finally, both the wind stress $\vec{\tau}$ and the wind diameter $D$ are used to perform the EPV calculation for comparison.

$$V_e = \frac{\vec{\tau}}{\rho f D} \qquad (4)$$

A more comprehensive discussion about the EPV and thermocline displacement is given in Section 4.

On the other hand, in this study, we also considered the forcing time of typhoons (Table 1), i.e., the typhoons' max wind blowing time $T_b$ in the region (see Appendix for details). Additionally, the adjustment times were calculated for the typhoons passing the study area (Table 1). The adjustment times were considered by assuming that the upwelling is weak at the beginning of wind forcing and that the potential upwelling velocity is only valid after a long enough time of the geostrophic adjustment process. According to the geostrophic adjustment theory [Gill 1982], such an adjustment time $T_a$ requires at least a $T_a = 1/f$ to be well established.

The altimeter data were derived from multi-sensors. These sensors included Jason-1, TOPEX/POSEIDON, GFO (Geosat Follow-On), ERS-2 and Envisat. Data were produced and distributed by AVSIO (Archiving, Validation and Interpretation of Satellite Oceanographic data). Near-real-time merged (TOPEX/POSEIDON or Jason-1 + ERS-1/2 or Envisat) sea surface height anomaly (SSHA) data, which are high resolutions of the 1/4°×1/4° Mercator grid, are available at www.aviso.oceanobs.com. The geostrophic currents, derived from SSHA with a resolution of 1/4°×1/4°, is also available at www.aviso.oceanobs.com.

## 3. Results

*3.1 Impact of Hagibis on Chl-a concentration*

Hagibis (2007) was a very weak typhoon and had a very special track (Fig 1a). It was generated from a tropical depression east of the Philippines on November 19, 2007, and it was upgraded to a tropical storm soon after. On November 21, the storm strengthened to typhoon status (central pressure of 963 mb and maximum wind speed of 41 m/s) and was named Hagibis. This category-1 typhoon moved slowly in the SCS, with a horizontal "V"-type track, and it passed over the middle of the SCS twice. It stayed in the SCS for more than five days.

Although Hagibis (2007) was very weak, it immediately had great impacts on the sea surface temperature. Before the typhoon's passage, there was warm water (with SST>26°C) in the middle of the SCS (Fig 2a). After the typhoon's passage, the strong winds caused by Hagibis (2007) caused sea surface cooling in a wide area range, with a maximum temperature cooling of -7°C beneath the former typhoon track (Fig 2b). This max sea surface cooling on November 25 lagged the typhoon's passage for about three days. It is noted that the cooling lasted for about three weeks (Fig 2c, d), much longer than in previous studies [Zhao et al. 2007, Gierach & Subrahmanyam 2008].

Meanwhile, the sea surface Chl-a concentration was significantly enhanced during this period. The pre-typhoon Chl-a concentrations were as low as 0.1-0.2 mg/m$^3$ in the SCS (Fig 3a). It is notable that the post-typhoon Chl-a concentrations increased as high as 5.0 mg/m$^3$ (increasing more than 20 times) in some places during the period from November 11 to November 30 (Fig 3b). This strong Chl-a concentration enhancement lasted for more than three weeks, and the Chl-a concentrations were still very high (>3.0 mg/m$^3$) from December 1 to December 15 (Fig 3c). The Chl-a concentration patterns (Fig 3b, c) coincided well with those of the sea surface temperature fields (Fig 2b, c). Additionally, this Chl-a enhancement should be due to local processes (vertical mixing, entrainment or upwelling) rather than non-local processes (e.g. horizontal convection from the coastal region [Zhao & Tang 2007]) because there was a low Chl-a concentration band separating the offshore high from the costal high. Finally, the blooming died out a month after the typhoon's passage (Fig 3d).

To show how big of an impact Typhoon Hagibis had made, the pixel distributions of Chl-a concentrations (in the study area) are shown in Fig 4a, including the cases of both pre-typhoon (dashed) and post-typhoon (solid). The Chl-a concentration peak increased from 0.11 mg/m$^3$ (pre-typhoon) to 0.21 mg/m$^3$ (post-typhoon). Additionally, there were significantly higher pixel values (>1.0 mg/m$^3$) after the typhoon's passage. The area average of Chl-a concentration is depicted in Fig 4b, where the pre-typhoon and post-typhoon Chl-a concentrations are 0.14 mg/m$^3$ and 0.74 mg/m$^3$, respectively. Hagibis (2007) had great impacts on the Chl-a concentration in the middle of the SCS for about one month. It is clear that the long-term phytoplankton blooming was due to upwelling of nutrients but was instead due to upwelling of Chl-a [Walker et al. 2005].

*3.2 Contribution to the climatology distribution in the middle of the SCS*

Moreover, the above-mentioned enhancement of Chl-a concentration was so distinctive that it also contributed to the climatology distribution in the middle of the SCS. To illustrate this, the Chl-a mean of ten years (1997.09-2007.09) in the SCS is depicted in Figure 5. The Chl-a concentration is relatively higher in the coastal region and islands compared to the off-shore region. The Chl-a concentration also has notable seasonal variations. In the summer, there is a high Chl-a concentration offshore of Vietnam (Fig 5b) because the southwest-northeast monsoon winds are roughly parallel to the coastline southeast of Vietnam and favorable wind-direction and strong wind magnitudes can lead to coastal upwelling in coastal areas and Ekman upwelling in offshore areas [Zhao & Tang 2007].

The impact of Hagibis on the Chl-a concentration in the middle of the SCS (the box in Fig 1) can be seen from Fig 6. In Fig 6a, the monthly mean of the Chl-a concentration within the middle of the SCS during the

ten-year period (1997.09-2007.09) is depicted. Before the year 2007, the Chl-a concentration exhibited similar seasonal variations; the concentration was lowest in May (0.1 mg/m$^3$) and highest in August (0.25 mg/m$^3$), which can also be seen in Fig 6b (solid curve). Additionally, the annual Chl-a concentration is about 1.78 mg/m$^3$ by integration of the monthly means within the region. However, the Chl-a concentrations in November and December of 2007 (0.467 mg/m$^3$ and 0.416 mg/m$^3$) were significantly higher than before (Fig 6b), when Typhoon Hagibis (2007) passed over the SCS. Thus, the monthly means of Chl-a concentrations notably increased from 0.157 mg/m$^3$ and 0.184 mg/m$^3$ to 0.188 mg/m$^3$ and 0.207 mg/m$^3$ due to the anomaly in the two months (dashed curve in Fig 6a). It is estimated that the Chl-a increase (0.55 mg/m$^3$ in two months) induced by Typhoon Hagibis (2007) accounts for about 30% of the annual Chl-a concentration (1.78 mg/m$^3$) and 2.8% of the total Chl-a concentration (19.6 mg/m$^3$) in the middle of the SCS.

In other words, Typhoon Hagibis (2007) had a great impact on the Chl-a concentration in the middle of the SCS. The amount of Chl-a enhancement even accounted for about 30% of the annual Chl-a concentration and 2.8% of the total Chl-a concentration.

### 3.3 Physical mechanism

As mentioned above, Typhoon Hagibis (2007) had great impacts on the Chl-a concentration in the middle of the SCS. The SCS is not only a region of typhoon passage, but it is also an area of significant tropical cyclone genesis [Wang et al. 2007]. It was found that there were 28 typhoons passing through the middle of the SCS during 1997-2007 (Fig 1b), and 11 of them were still at typhoon status (Table 1). However, other typhoons, even though much stronger, had little impact on the Chl-a enhancement in the middle of the SCS. As the surface phytoplankton blooming is associated with the local upwelling and the thermocline displacement by typhoons, it seems that Typhoon Hagibis (2007) induced the strong upwelling and the large thermocline displacement.

We first considered the potential typhoon-induced Ekman pumping. Figure 7 depicts the potential EPV fields derived from the wind fields during Typhoon Hagibis's passage. The EPV maximum values were 30~50×10$^{-4}$ m/s on the evening of November 21 (Fig 7a) and 50~60×10$^{-4}$ m/s on the night of November 22 (Fig. 7b) when Hagibis crept westward to the coastline southeast of Vietnam. Comparing the present result with those of previous studies [Lin et al. 2003, Shi & Wang 2007, Gierach & Subrahmanyam 2008, Shang et al. 2008], it is concluded that the EPV was quite larger in this case (Table 2). The reason why the EPV was quite large is mainly due to the fact that the location of the typhoon (at 10°N) is close to equator, and the Coriolis parameter is about half as small compared with that in the other typhoons at 20°N. Thus, the potential EPV calculated from Eq. (1) should be twice as large as that due to the typhoon with the same strength at a higher latitude.

Additionally, the longer forcing time of Hagibis (2007) played the most important role in the establishment of the potential EPV. According to the geostrophic adjustment theory, such an upwelling process requires a geostrophic adjustment time of at least $T_a = 1/f$ to be well established. If the typhoon's forcing time is not long enough, the upwelling velocity should be much less than the potential EPV. In Table 1, both the forcing time and the adjustment time for the typhoons are shown. For most of the typhoons, the forcing time is significantly smaller than the adjustment time, especially for the fast-moving typhoons; thus, the potential EPV cannot be achieved for the upwelling velocity. Because Typhoon Hagibis had an extremely longer forcing time (>82 hours) than the required time (~63 hours) the strong potential EPV was nearly reached. The larger upwelling velocity and the longer forcing time made the thermocline displacement much larger, finally leading to great biophysical responses.

Finally, the vertical mixing and the upwelling by Hagibis also induced a cyclonic eddy, as the SSHA decreased right beneath the typhoon track (Fig 8a-d). Meanwhile, the mixed layer depth (MLD) around the eddy deepened about 25 m due to stirring from the temperature profiles of Argo float 5900059 (Fig 8e), which implies that the MLD deepening in the eddy center might be much larger. Comparing Fig 8 with Fig 7, the center of the cyclonic eddy is found to coincide with the position of the SST cooling center, which implies that both phenomena were due to the same local mixing and upwelling processes. Such a cyclonic eddy also induced a strong cyclonic flow. The enhanced cyclonic eddy induced a strong geostrophic current along the coast (~1.0 m/s), compared with the weak surface flow (~0.4 m/s) before the typhoon. It was found here that the Argo float track was along the geostrophic current (Fig 8a-d) and that the flow obtained from SSH agreed well with that obtained from Argo (Fig 8f). In consequence, it brought the high Chl-a waters to the south along the eddy. The high Chl-a concentration waters were located near the typhoon track at first (Fig 3b), and they were then transported along the coast to the south by surface flow (Fig 3c). According to Eq.(3), the thermocline displacement was about 100 m, which is quite larger than that in other cases (Table 2). The cyclonic eddy was not stable, and it then broke into two small eddies from November 29 to December 11. The bigger eddy went northward, and the smaller eddy went southward.

## 4. Discussion

The above investigation shows that forcing time might be the most important factor for ocean responses to typhoons. As shown in Table 1, there are four typhoons with forcing times larger than 40 hours, besides Typhoon Hagibis (2007). Two of them, Lingling (2001) and Kai-Tak (2005), have been investigated in the literatures [Shang et al. 2008, Zheng & Tang 2007, Zhao et al. 2008]. Both had notable impacts on the regional Chl-a increase. The last typhoon, Chanchu (2006), not only made some temporarily notable (but relatively smaller in comparison with Typhoon Hagibis) impacts on the Chl-a concentration in the northern SCS (Fig 9), but it also had influences on the 2006 South China Sea summer monsoon onset [Mao and Wu, 2008]. Other typhoons, regardless of whether they were higher in intensity, had smaller impacts on the ocean responses and have not been investigated in the literature.

Compared with forcing time, the typhoon intensity might be less important. This can also be understood from the reduced drag coefficient for high wind speeds in tropical cyclones [Powell et al. 2003, Jarosz et al. 2007]. Thus, the wind stress under high wind speeds (>50 m/s) is only slightly larger than that under relatively lower wind speeds (30~40 m/s).

Although the pre-existing oceanic conditions (cold core eddies) also played little roles in this case, they are often important in the upper ocean's responses [Walker et al. 2005, Shi & Wang 2007, Zheng et al. 2008, Sun et al. 2009]. In fact, in some cases, the cyclonic eddies became the major indices for sea surface cooling [Zheng et al. 2008, Wada et al. 2009]. In this case, one big, cold core eddy was even generated due to the typhoon forcing (Fig 8b). This might be one of the cold eddy genesis mechanisms.

The advection of Chl-a from one place to another would also cause a temporal Chl-a concentration increase [Zhao & Tang 2007, Yang et al. 2010]. In this case, the original Chl-a concentration enhancement was located beneath the typhoon track. Then, such Chl-a was advected to the south along the eddy edge for about 300 km (even at the left side of typhoon track). This implies that the impact of typhoons on surface Chl-a by advection could be much wider than previously known.

In the literature and in the above investigation, the EPV is always calculated by Eq. (2). We also used the "best track datasets," new drag law [Jarosz et al. 2007] and Eq. (4) to calculate the EPV (Table 2). It was found that the EPV induced by Typhoon Hagibis (2007) is still the largest, although the EPVs calculated by Eq. (4) are different from those using Eq. (2).

As we know, the enhancement of the Chl-a concentration depends on the Ekman pumping of subsurface nutrients and hence on the typhoon-induced wind stress and the forcing time. Thus, the physical formula is important for relating the thermocline displacement in the upper ocean to a typhoon's parameters, e.g., a typhoon's intensity, wind stress and moving speed [[Price 1981]. For this purpose, the thermocline displacement is estimated to be directly proportional to the typhoon-induced surface wind stress and inversely proportional to the typhoon's moving speed [Price 1994], as seen in Eq. (3), which can be understood from the conception of translation time [Lin et al. 2008]. Such conception assumes that the typhoon-induced response is directly proportional to the typhoon's diameter $D$ and inversely proportional to the typhoon's moving speed $U_T$. Thus, the thermocline displacement is estimated to be the product of potential EPV and translation time,

1  $$\Delta\eta = \frac{curl(\vec{\tau})}{\rho f} \times \frac{D}{U_T} \simeq \frac{\vec{\tau}/D}{\rho f} \times \frac{D}{U_T} = \frac{\vec{\tau}}{\rho f U_T}, \quad (5)$$

2  which is exactly the same with Eq. (3).

3  However, such estimation would overestimate the typhoon's forcing, as the forcing times of the typhoons (e.g. typhoon in Table 1) are often not long enough for the potential EPV to be achieved. For the fast-moving typhoons, the wind-induced upwelling should be very weak, and the thermocline displacement might be inversely proportional to the typhoon's moving speed square. Nevertheless, how to make a more accurate estimation is still an open problem for further studies. The in situ observations [Dickey et al. 1998, Zedler et al. 2002, Black & Dickey 2008] would be very helpful for such investigations.

Overall, the reasons why Typhoon Hagibis (2007) made such a great impact on the climatology and Chl-a concentration in the middle of the SCS are due to the location of the typhoon and the typhoon's long-term forcing time (Fig 10). Thus, the potentially strong upwelling was realistically established. Besides, the advection can take the extra Chl-a 300 km away from the typhoon track. Compared with those parameters, the intensity of the typhoon, the wind speed and the upwelling of Chl-a played minor roles in this case.

## 5. Conclusion

In 2007, Typhoon Hagibis (category-1) made a great impact on the Chl-a concentration, although there were more than 20 typhoons passing over the region. This slowly moving typhoon, with a horizontal "V"-type track, stayed in the SCS for more than five days. The long-term forcing time (>82 hours) due to the typhoon induced the strong upwelling and caused a significant blooming of phytoplankton, which accounted for 30% and 2.8% of the annual and total Chl-a concentrations (1997-2007), respectively, in the middle of the South China Sea (SCS). Compared with those parameters, the intensity of the typhoon, the wind speed and the upwelling of Chl-a played minor roles in this case. It is also argued that Price's formula overestimates the typhoon-forced upwelling and that a more accurate formula is needed.

**Acknowledgements:**

We thank four anonymous reviewers for their constructive suggestions. This work is supported by the National Basic Research Program of China (No. 2007CB816004), the Knowledge Innovation Program of the Chinese Academy of Sciences (Nos. KZCX2-YW-QN514 and KZCX2-YW-226), the National Foundation of Natural Science (Nos. 40705027 and 40730950), and the Open Fund of State Key Laboratory of Satellite Ocean Environment Dynamics (No. SOED0902). We thank Shanghai Typhoon Institute (STI) of China Meteorological Administration (CMA) for providing typhoon track data, China Argo Real-time Data center for the Argo float profiles, NASA's Ocean Color Working Group for providing Modis and SeaWiFS data, AVISO for SSHA data and Remote Sensing Systems for TMI/AMSR-E SST data.

**Appendix**

The forcing time of typhoons is defined objectively in the following way. At first, we simply assume that the typhoon impact region is within the area where wind speed is over a critical value $U_c$ (e.g. $U_c$ =17 m/s) along its track. Secondly, for each point, the blowing time of the point is defined as the total time for which the wind speed is over $U_c$. Additionally, the max total blowing time is defined as the forcing time in this region. Numerical implementation would require that every six-hour track is interpolated as the short-time (e.g. half-hour) track. Figure A1a shows the blowing time calculated by this method, where the contour curve labeled '40' means that the water within this region has a forcing time larger than 40 hours. The forcing time in this region is calculated at about 82 hours with an error less than 1 hour (Figure A1a).

We can also roughly estimate the forcing time by following the conclusion drawn by Hanshaw et al. (2008). It is simply assumed that the typhoon impact region is within the area where the distance between the point and typhoon track is less than 200 km. Then, we draw a sample circle box with a diameter of 200 km and move this sample box to ensure that it completely covers the tracks (Figure A1b). After summing the forcing times by accounting for the tracks in the box, the forcing time in this region was calculated as ~102 hours with an error less than 3 hours (half of the original typhoon track time interval).

Figure A1c shows the forcing time using a variable radius in comparison with that using a fixed 200-km radius. As the typical forcing radius is about 150 km for a category-1 typhoon, a fixed 200-km radius would lead to overestimation for weak typhoons like Hagibis (Fig A1b), while this fixed radius would lead to underestimation for strong typhoons.

**Reference:**

Babin SM, Carton JA, Dickey TD and Wiggert JD, (2004) Satellite evidence of hurricane-induced phytoplankton blooms in an oceanic desert, *J. Geophys. Res.*, 109, C03043, doi:10.1029/2003JC001938.

Black WJ, TD Dickey (2008) Observations and Analyses of Upper Ocean Responses to Tropical Storms and Hurricanes in the Vicinity of Bermuda, J Geophys Res, 113:C08009 doi:10.1029/2007JC004358.

Chang Y, Liao HT, Lee MA, Chan JW, Shieh WJ, Lee KT, Wang GH, Lan YC (2008) Multisatellite observation on upwelling after the passage of Typhoon Hai-Tang in the southern East China Sea. Geophys Res Lett 35: L03612, doi:10.1029/2007GL032858

Davis A, Yan XH (2004) Hurricane forcing on chlorophyll-a concentration off the northeast coast of the U.S. Geophys Res Lett 31:L17304, doi:10.1029/2004GL020668

Dickey T, Frye D, McNeil J, Manov D, Nelson N, Sigurdson D, Jannasch H, Siegel D, Michaels T, Johnson R (1998) Upper-ocean temperature response to Hurricane Felix as measured by the Bermuda Testbed Mooring. Mon Weather Rev, 126, 1195-1201.

Garratt JR (1977) Review of drag coefficients over oceans and continents. Mon Weather Rev 105:915-929

Gierach MM, Subrahmanyam B (2008) Biophysical responses of the upper ocean to major Gulf of Mexico hurricanes in 2005. J Geophys Res C 113:C04029, doi:10.1029/2007JC004419

Gill AE (1982) Atmosphere-Ocean Dynamics. Academic Press, London

Gregg WW, Casey NW, McClain CR (2005) Recent trends in global ocean chlorophyll. Geophys Res Lett

32:L03606, doi:10.1029/2004GL021808

Lin I, Liu WT, Wu CC, Wong GTF, Hu C, Chen Z, Liang WD, Yang Y, Liu KK (2003) New evidence for enhanced ocean primary production triggered by tropical cyclone. Geophys Res Lett 30(13):1718, doi:10.1029/2003GL017141

Lin II, Wu CC, Pun IF, Ko DS (2008) Upper ocean thermal structure and the western North Pacific category-5 typhoons. Part I: Ocean features and category-5 typhoon's intensification. Mon Weather Rev 136:3288– 3306

Hanshaw MN, Lozier MS, Palter JB (2008) Integrated impact of tropical cyclones on sea surface chlorophyll in the North Atlantic. Geophys Res Lett 35:L01601, doi:10.1029/2007GL031862

Jarosz E, Mitchell DA, Wang DW, Teague WJ (2007) Bottom-Up Determination of Air-Sea Momentum Exchange Under a Major Tropical Cyclone. Science 315:1707, doi: 10.1126/science.1136466

Mao J, Wu G (2008) Influences of Typhoon Chanchu on the 2006 South China Sea summer monsoon onset Geophys Res Lett: L12809, doi: 10.1029/2008GL033810

McClain CR, Signorini SR, Christian JR (2004) Subtropical gyre variability observed by ocean-color satellites. Deep-Sea Res Part II 51: 281-301

McClain CR (2009) A Decade of Satellite Ocean Color Observations. Annu Rev Marine Sci 1:19-42

Powell, Mark D., P. J. Vickery, and T. A. Reinhold (2003) Reduced drag coefficient for high wind speeds in tropical cyclones. Nature, 422, 279-283

Price JF, Sanford TB, Forristall GZ (1994) Forced stage response to a moving hurricane. J Phys Oceanogr 24: 233-260

Shang S, Li L, Sun F, Wu J, Hu C, Chen D, Ning X, Qiu Y, Zhang C, and Shang S (2008) Changes of temperature and bio-optical properties in the South China Sea in response to Typhoon Lingling, 2001. Geophys Res Lett 35:L10602, doi:10.1029/ 2008GL033502

Shi W, Wang M (2007) Observations of a Hurricane Katrinainduced phytoplankton bloom in the Gulf of

Mexico. Geophys Res Lett 34:L11607, doi:10.1029/2007GL029724

Siswanto E, Ishizaka J, Morimoto A, Tanaka K, Okamura K, Kristijono A, Saino T (2008) Ocean physical and biogeochemical responses to the passage of Typhoon Meari in the East China Sea observed from Argo float and multiplatform satellites. Geophys Res Lett 35:L15604, doi:10.1029/2008GL035040

Subrahmanyam B, Rao KH, Rao SN, Murty VSN, Sharp RJ (2002) Influence of a tropical cyclone on chlorophyll-a concentration in the Arabian Sea. Geophys Res Lett 29:2065, doi:10.1029/2002GL015892

Sun L, Yang YJ, Fu YF (2009) Impacts of Typhoons on the Kuroshio Large Meander: Observation Evidences. Atmos Ocean Sci Lett 2:40-45

Walker ND, Leben RR, Balasubramanian S (2005) Hurricane-forced upwelling and chlorophyll a enhancement within cold-core cyclones in the Gulf of Mexico. Geophys Res Lett 32:L18610, doi:10.1029/2005GL023716

Wada A, Sato K, Usui N, Kawai Y (2009) Comment on "Importance of pre-existing oceanic conditions to upper ocean response induced by Super Typhoon Hai-Tang" by Z.-W. Zheng, C.-R. Ho, and N.-J. Kuo, Geophys Res Lett, 36, L09603, doi:10.1029/2008GL036890.

Wang GH, Chen D, Su JL (2008) Winter Eddy Genesis in the Eastern South China Sea due to Orographic Wind Jets. J Phys Oceanogr 38:726-732

Wang GH, Chen D, Chu P (2003) Mesoscale eddies in the South China Sea observed with altimeter data. Geophys Res Lett 30:2121, doi:10.1029/2003GL018532

Yang YJ, Sun L, Liu Q, Xian T, Fu YF (2010) The Biophysical Responses of the upper ocean to the typhoons Namtheun and Malou in 2004. Int J Rem Sen in press

Zedler SE, Dickey TD, Doney SC, Price JF, Yu X, Mellor GL, (2002) Analyses and simulations of the upper ocean's response to Hurricane Felix at the Bermuda Test bed Mooring site: 13-23 August 1995, J Geophys Res, 107 (12), 25-1. doi:10.1029/2001JC00969

Zhao H, Tang DL (2007) Effect of 1998 El Nino on the distribution of phytoplankton in the South China Sea.

J Geophys Res 112:C02017, doi:10.1029/2006JC003536

Zhao H, Tang DL, Wang Y (2008) Comparison of phytoplankton blooms triggered by two typhoons with different intensities and translation speeds in the South China Sea. Mar Ecol Prog Ser 365:57-65, doi:10.3354/meps07488

Zheng GM, Tang DL (2007) Offshore and nearshore chlorophyll increases induced by typhoon winds and subsequent terrestrial rainwater runoff. Mar Ecol Prog Ser 333:61-74

Zheng ZW, Ho CR, Kuo NJ (2008) The importance of pre-existing oceanic conditions to upper ocean response induced by Super Typhoon Hai-Tang. Geophys Res Lett 35: L20603, doi:10.1029/2008GL035524

Table 1. The typhoons passing over the study area, where MWS/BWS, CT, MS, EPV, AT and FT represent max wind speed/best track wind speed, category of typhoon, moving speed, Ekman pumping velocity $V_e$, adjustment time $T_a$ and forcing time $T_b$, respectively. The superscripts "a" and "b" represent the values calculated according to Eq. (2) and Eq. (4), respectively.

| Typhoon (month, year) | MWS/BWS (m/s) | CT | MS (m/s) | Latitude (°N) | Max EPV ($10^{-4}$m/s) | AT (h) | FT (h) |
|---|---|---|---|---|---|---|---|
| Faith (Dec. 1998) | 30/46.2 | 2 | 4.3 | 10-12 | N.A.[a] /N.A.[b] | 63 | N.A. |
| Lingling (Nov. 2001) | 50/59 | 4 | 4.6 | 13 | 45.2[a]/ 21.6[b] | 53 | 40 |
| Nepartak (Nov. 2003) | 30/38.5 | 1 | 4.6 | 13-15 | 7.8[a] / 17.8[b] | 50 | 20 |
| Muifa (Nov. 2004) | 30/46.2 | 2 | 3.1 | 9-12 | 10.7[a] /N.A.[b] | 69 | 26 |
| Chathu (Jun. 2004) | 33/38.5 | 1 | 6.6 | 12-14 | 12.9[a] /N.A.[b] | 53 | 15 |
| Kai-tak (Oct. 2005) | 40/43.6 | 2 | 2.2 | 12-15 | 15.7[a] /17.9[b] | 53 | 52 |
| Chanchu (May. 2006) | 50/64 | 4 | 2.0 | 14-15 | 21.5[a] /32.9[b] | 50 | 44 |
| Chebi (Nov. 2006) | 30/36 | 1 | 4.1 | 15 | 2.9[a] /16.8[b] | 45 | 16 |
| Durian (Dec. 2006) | 40/46.2 | 2 | 4.2 | 11-14 | 38.7[a] /19.6[b] | 53 | 26 |
| Utor (Dec. 2006) | 45/41 | 1 | 4.5 | 14-15 | 14.8[a] /26.1[b] | 50 | 18 |
| Hagibis (Nov. 2007) | 35/41 | 1 | 2.5 | 10 -12 | 48.1[a] /34.6[b] | 63 | 82 |

Table 2: The typhoons and their locations (latitude), MWS (max wind speed), CT (category of typhoon), MS (translation speed), EPVs (Ekman pumping velocity using Eq. (2) and QuikSCAT data), dMLDs (difference of mixed layer depths)/TDs (thermocline displacement using Eq. (3) and QuikSCAT data) and references.

| Typhoon (year) | Latitude | MWS (m/s) | CT | MS (m/s) | max EPV ($10^{-4}$m/s) | dMLD/ TD(m) | References |
|---|---|---|---|---|---|---|---|
| Kai-Tak (2000) | 20°N | 48 | 2 | 4.5 | 20 | 21/90. | Lin et al., 2003 |
| Lingling (2001) | 14°N | 59 | 4 | 5 | N.A. | 20/100 | Shang et al., 2008 |
| Hai-Tang (2005) | 25°N | 70 | 5 | 11 | 0.7 | N.A. | Chang et al., 2008 |
| Kai-Tak(2005) | 13°N | 46 | 2 | 2.9 | 1.7 | 23.5/ N.A. | Zheng & Tang, 2007 |
| Katrinia (2005) | 24°N | 77 | 5 | 3.4 | 20 | N.A. /80. | Shi & Wang, 2007 |
| Katrinia (2005) | 24°N | 77 | 5 | 3.4 | 5 | N.A./ 63 | Gierach and Subrahmanyam, 2008 |
| Rita (2005) | 27°N | 80 | 5 | 4.5 | 12 | N.A./ 87 | |
| Wilma (2005) | 21°N | 82 | 5 | 3.6 | 7 | N.A./ 45 | |
| Hagibis (2007) | 10°N | 41 | 1 | 2.5 | 48.1 | 25/105 | Present |

Figure 1 (a) Track of Typhoon Hagibis (2007) in the study area. Typhoon center positions every six hours are indicated. (b) The typhoon tracks in the South China Sea during 1997-2007.

Figure 2 The SST cooling during the typhoon's passage, with a maximum cooling temperature of -7°C beneath the former typhoon track.

Figure 3 The strong chlorophyll-a enhancement (lasted for more than three weeks) during the typhoon's passage.

Figure 4 (a) The distributions of chlorophyll-a (Chl-a) pixels pre- (dashed) and post- (solid) typhoon. The Chl-a concentration peak increased from 0.11 mg/m$^3$ (pre-typhoon) to 0.21 mg/m$^3$ (post-typhoon). (b) The time series of the area average of Chl-a, where typhoon-induced blooming lasted more than three weeks.

Figure 5 The seasonal means of chlorophyll-a concentrations in the South China Sea. The Chl-a concentration is relatively higher in the coastal region and islands than in the off-shore region.

Figure 6 (a) The monthly means of chlorophyll-a pre- (dashed) and post- (solid) typhoon, and the number of tropical cyclones that passed over the study area. (b) The time series of average chlorophyll-a concentrations and tropical cyclones. The impact of Hagibis (2007) on the chlorophyll-a concentration is notable.

Figure 7 The Ekman pumping velocity (EPV) during the typhoon's passage. The max EPV ($50\sim60\times10^{-4}$ m/s) was quite large in this case.

Figure 8 (a)-(d) The sea surface height anomaly (SSHA) during the typhoon's passage, where the black points mark the positions of Argo float 5900059. The enhanced cyclonic eddy induced a strong southward surface flow along the coast, while the surface flow was weak before the typhoon passed. (e) The temperature profiles before and after the typhoon. The mixed layer depth deepened about 25 m after the typhoon. (f) The surface flow speeds detected by the Argo float and geostrophic currents both have good correlation with each other.

Figure 9 The ocean surface chlorophyll-a concentration in response to Typhoon Chanchu (2006) was notable but relatively smaller than that in response to Typhoon Hagibis (2007).

Figure 10 The diagram of ocean responses due to Typhoon Hagibis (2007).

Figure A1 (a) Blowing time along the track by a variable forcing radius at wind speed over 15 m/s, with a maximum blowing time (forcing time) of ~82 hours. The contour curve labeled '40' indicates a long-term forcing region having a forcing time longer than 40 hours. (b) Forcing time (~102 hours) calculated by a moving circle sample box with a radius of 200 km. (c) Forcing times by variable radii vs. those by a fixed radius of 200 km for all the typhoons.

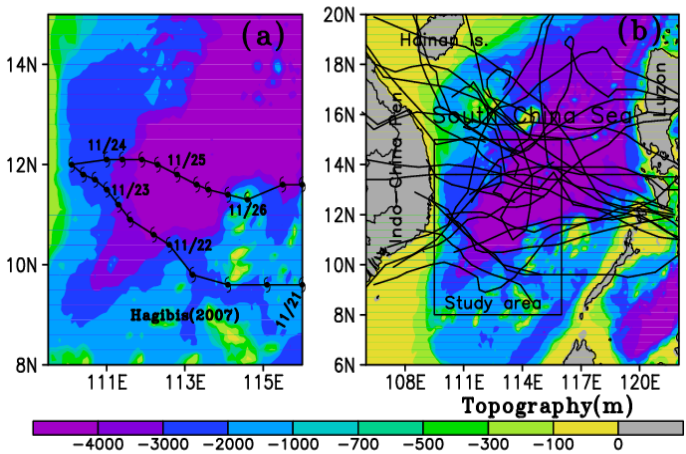

Figure 1 (a) Track of Typhoon Hagibis (2007) in the study area. Typhoon center positions every six hours are indicated. (b) The typhoon tracks in the South China Sea during 1997-2007.

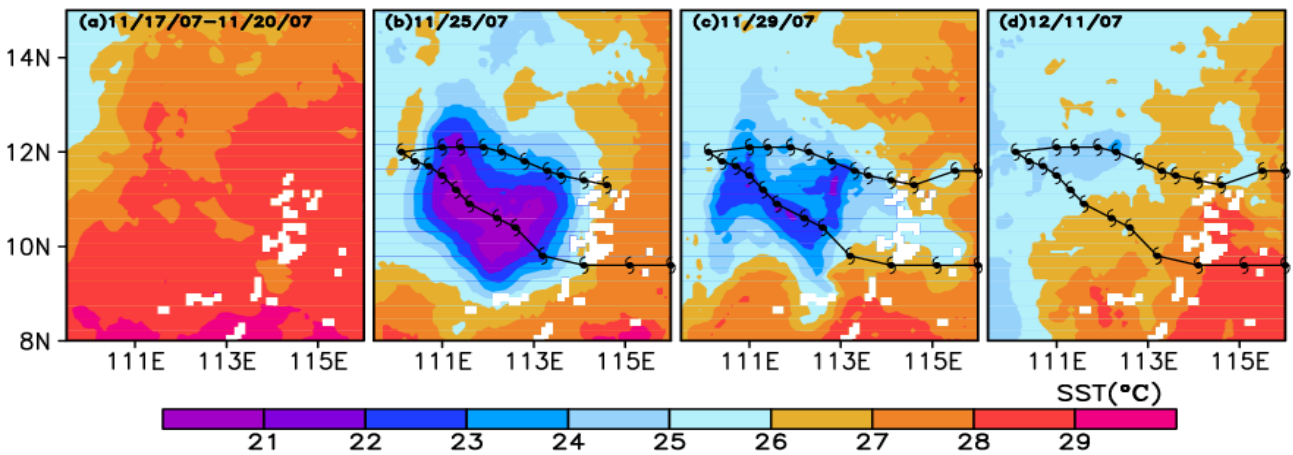

Figure 2 The SST cooling during the typhoon's passage, with a maximum cooling temperature of -7°C beneath the former typhoon track.

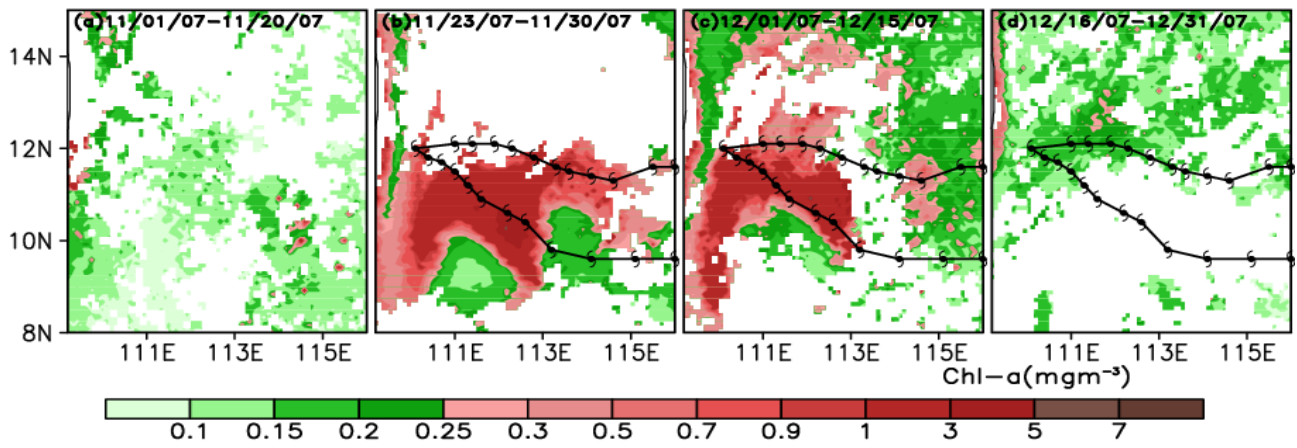

Figure 3 The strong chlorophyll-a enhancement (lasted for more than three weeks) during the typhoon's passage.

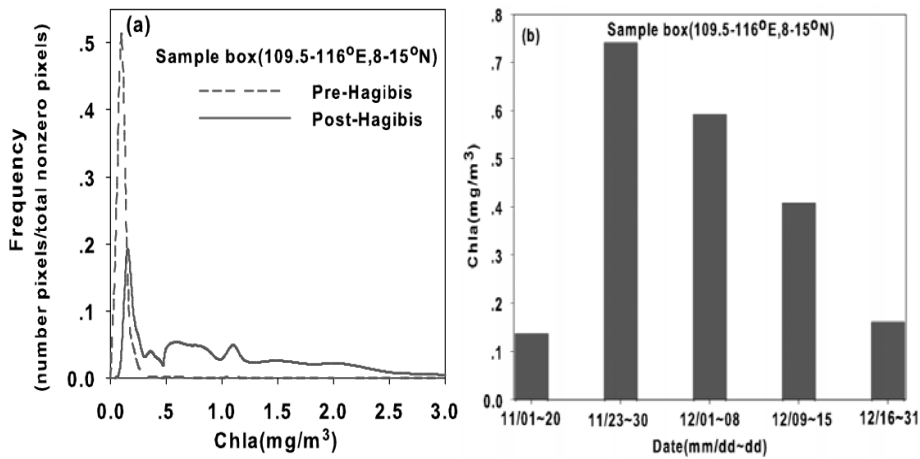

Figure 4 (a) The distributions of chlorophyll-a (Chl-a) pixels pre- (dashed) and post- (solid) typhoon. The Chl-a concentration peak increased from 0.11 mg/m$^3$ (pre-typhoon) to 0.21 mg/m$^3$ (post-typhoon). (b) The time series of the area average of Chl-a, where typhoon-induced blooming lasted more than three weeks.

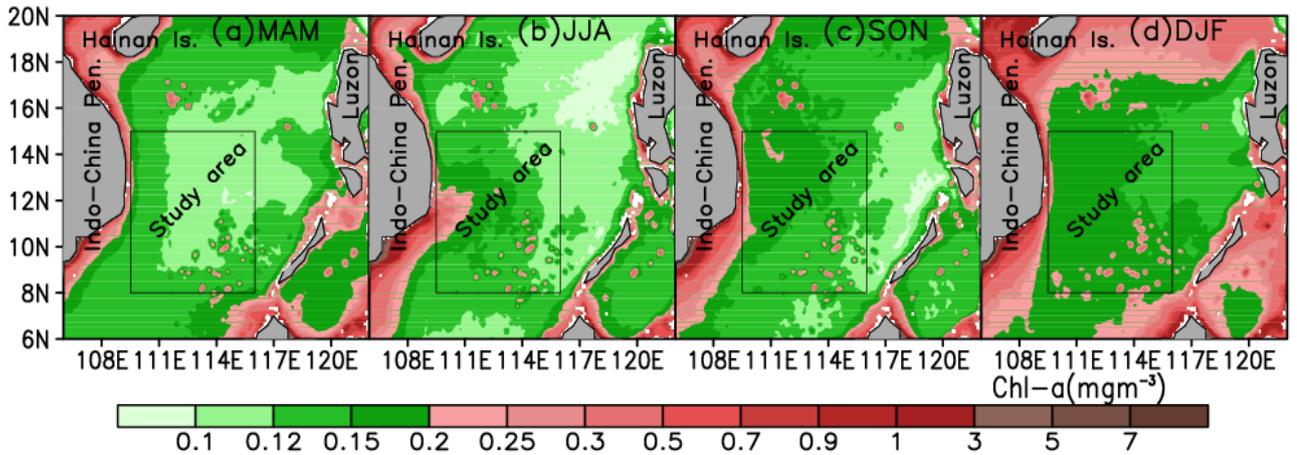

Figure 5 The seasonal means of chlorophyll-a concentrations in the South China Sea. The Chl-a concentration is relatively higher in the coastal region and islands than in the off-shore region.

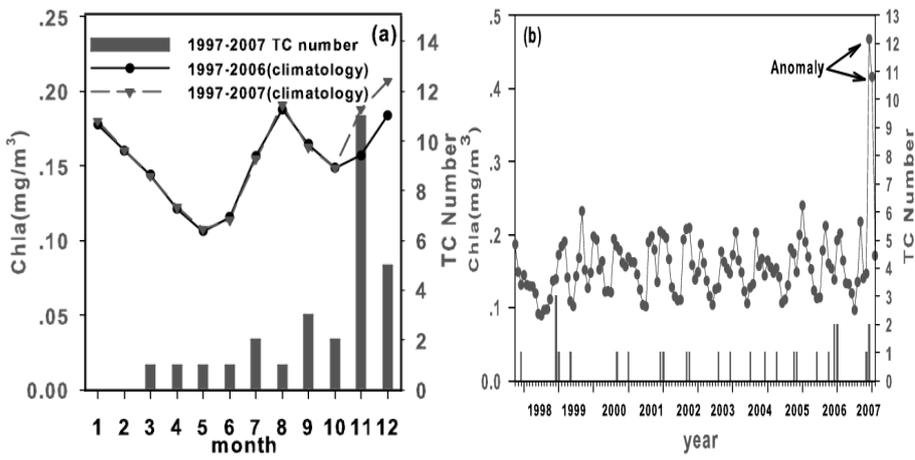

Figure 6 (a) The monthly means of chlorophyll-a pre- (dashed) and post- (solid) typhoon, and the number of tropical cyclones that passed over the study area. (b) The time series of average chlorophyll-a concentrations and tropical cyclones. The impact of Hagibis (2007) on the chlorophyll-a concentration is notable.

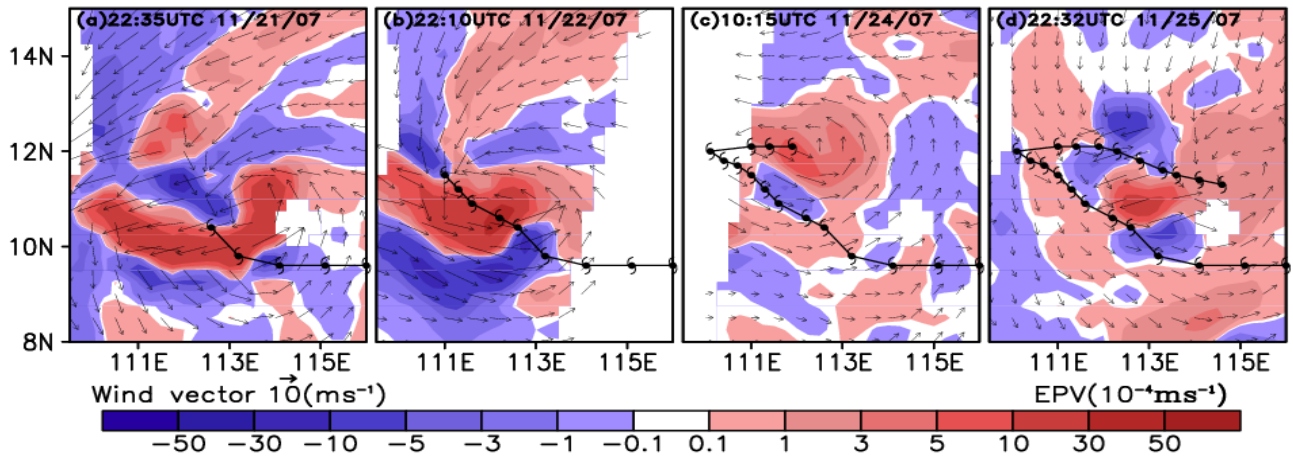

Figure 7 The Ekman pumping velocity (EPV) during the typhoon's passage. The max EPV ($50\sim60\times10^{-4}$ m/s) was quite large in this case.

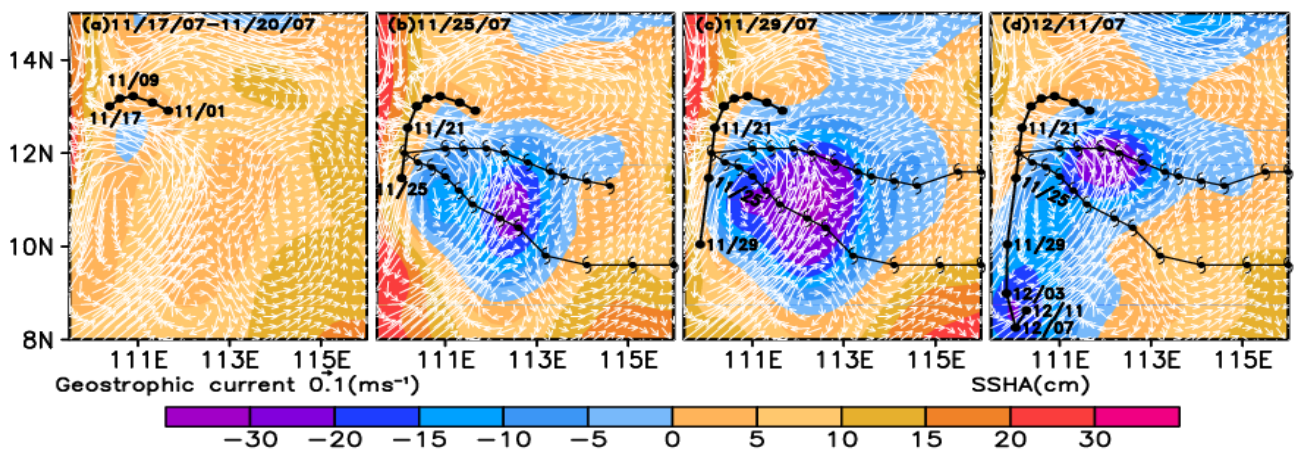

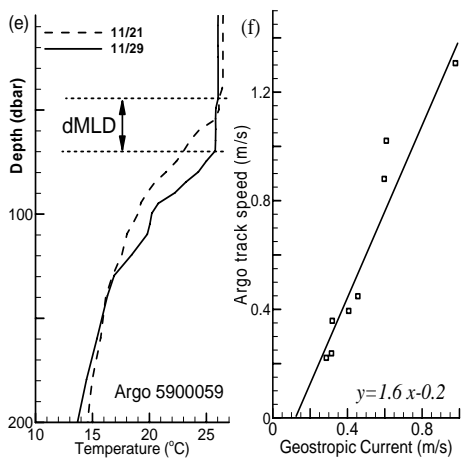

Figure 8 (a)-(d) The sea surface height anomaly (SSHA) during the typhoon's passage, where the black points mark the positions of Argo float 5900059. The enhanced cyclonic eddy induced a strong southward surface flow along the coast, while the surface flow was weak before the typhoon passed. (e) The temperature profiles before and after the typhoon. The mixed layer depth deepened about 25 m after the typhoon. (f) The surface flow speeds detected by the Argo float and geostrophic currents both have good correlation with each other.

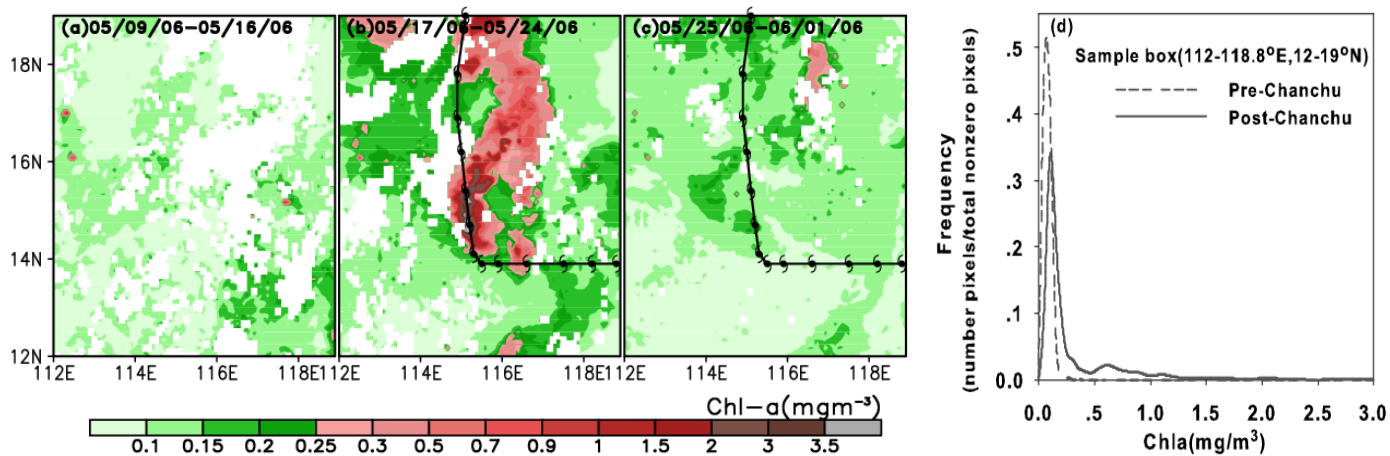

Figure 9 The ocean surface chlorophyll-a concentration in response to Typhoon Chanchu (2006) was notable but relatively smaller than that in response to Typhoon Hagibis (2007).

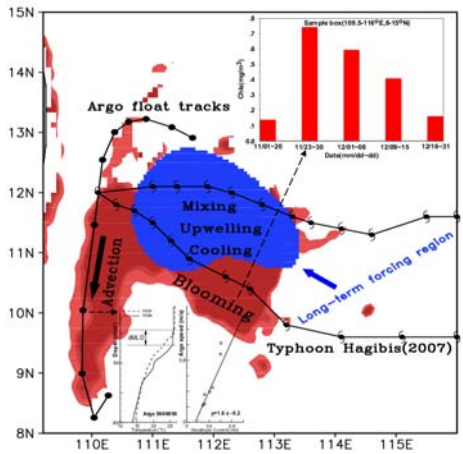

Figure 10 The diagram of ocean responses due to Typhoon Hagibis (2007).

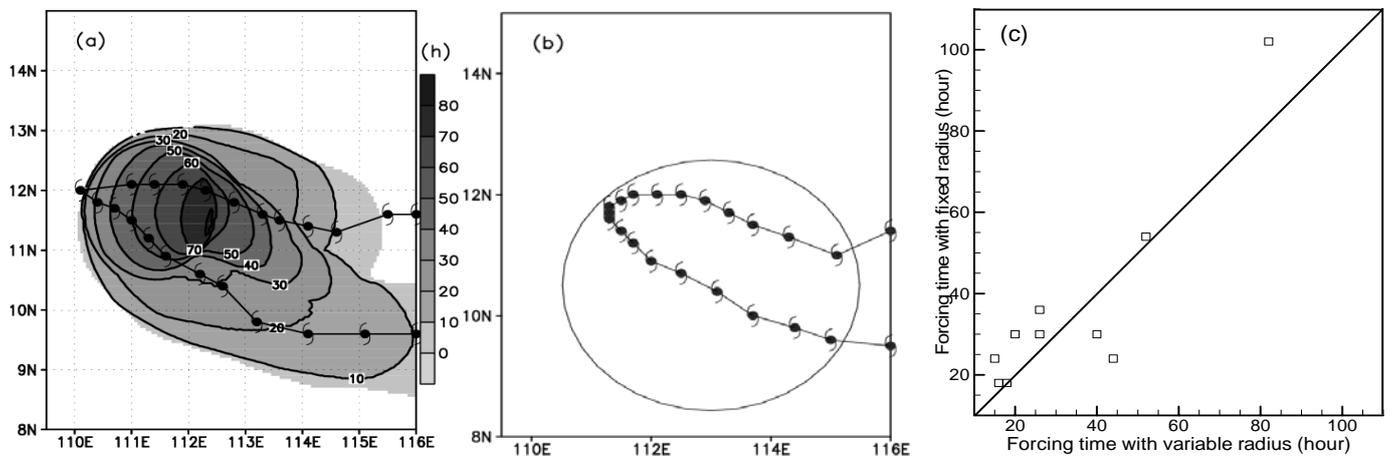

Figure A1 (a) Blowing time along the track by a variable forcing radius at wind speed over 17 m/s, with a maximum blowing time (forcing time) of ~82 hours. The contour curve labeled '40' indicates a long-term forcing region having a forcing time longer than 40 hours. (b) Forcing time (~102 hours) calculated by a moving circle sample box with a radius of 200 km. (c) Forcing times by variable radii vs. those by a fixed radius of 200 km for all the typhoons.